\documentclass[aps,prd,twocolumn,superscriptaddress,10pt,noshowpacs]{revtex4}  
\usepackage[active]{srcltx}
\usepackage{graphicx}
\usepackage[utf8]{inputenc}
\usepackage{amsmath}
\usepackage{xcolor}
\usepackage{times,txfonts}
\usepackage{float}%para fixar as figuras na posição desejada
\DeclareMathOperator{\csch}{csch}

\begin{document}

\title{Effects of Lorentz violation in the Bose-Einstein condensation}

\author{J. Furtado}
\affiliation{Centro de Ci\^{e}ncias e Tecnologia, Universidade Federal do Cariri, 57072-270, Juazeiro do Norte, Cear\'{a}, Brazil}
\email{job.furtado@ufca.edu.br}

\author{A. C. A. Ramos}
\affiliation{Centro de Ci\^{e}ncias e Tecnologia, Universidade Federal do Cariri, 57072-270, Juazeiro do Norte, Cear\'{a}, Brazil}
\email{antornio.ramos@ufca.edu.br}

\author{J. F. Assun\c{c}\~{a}o}
\affiliation{Departamento de F\'{i}sica, Universidade Regional do Cariri, 63105-000, Juazeiro do Norte, Cear\'{a}, Brazil}
%\email{jfassuncao@fis.ufal.br}

\date{\today}

\begin{abstract}
In this paper we study the corrections emergent from a Lorentz-violating CPT-odd extension of the complex scalar sector to the Bose-Einstein condensation and to the thermodynamics parameters. We initially discussed some features of the model to only then compute the corrections to the Bose-Einstein condensation. The calculations were done by computing the generating functional, from which we extract the thermodynamics parameters. We also obtained a Lorentz-violating correction for the critical temperature $T_c$ that sets the Bose-Einstein Condensation.
\end{abstract}

\maketitle

\section{Introduction}

In the last years possible extensions of the Standard Model (SM) have been studied and, in this context, Lorentz and CPT symmetries breaking are now considered as an essential topic of discussion \cite{Kos1, Kos2, Kos3, Kos4, Colladay:1996iz, Colladay:1998fq}. By introducing privileged directions in space-time the Lorentz symmetry is broken. Such privileged space-time directions are expressed through some additive terms which are proportional to small constant vectors or tensors. The Standard Model Extension (SME) \cite{Colladay:1996iz, Colladay:1998fq} is the most well established model that consider the effects of Lorentz and CPT symmetry violation.

Recently an extension of the scalar sector considering Lorentz-Violating effects was proposed by Kostelecky and Edwards \cite{Edwards:2018lsn}. Such model presents a general effective scalar field theory in any spacetime dimension containing explicit perturbative spin-independent Lorentz violating operators of arbitrary mass dimension. A topic of great importance addressed by this construction is the fact that the great majority of the fundamental particles of the SM have spin, being the Higgs boson the only example in the SM of a fundamental spinless particle. In spite of the minor role played by the scalar sector of QED (sQED), compared to strong interaction, in the description of coupling between mesons, it was argued \cite{Edwards:2019lfb} that a Lorentz-violating extension of sQED could be an effective way of treating tiny CPT deviations in neutral-mesons oscillations.

The finite temperature effects in the context of Lorentz symmetry violation have been extensively studied in the last years, specially in the context of radiative corrections \cite{Cervi:2001fg, Leite:2013pca, Leite:2011jg}, tree level scatterings \cite{Santos:2020dfn, Souza:2019zyt, Santos:2019hpw, Santos:2018jsi}, massless QED \cite{Brito:2009rp}, ambiguities in the Chern-Simons induction and non-analiticity \cite{Mariz:2005jh, Mariz:2014sia, Assuncao:2016fko}. The topic of Bose-Einstein condensation in the context of Lorentz-violating theories was considered in \cite{Casana:2011bv}, where it was computed the influence of a CPT-even term in the Bose-Einstein condensation. In this paper we are interested in the effects of a Lorentz-violating CPT-odd term in the finite temperature regime and particularly in the Bose-Einstein condensation. We perform the calculations by the explicit computation of generating functional, from which we obtain the thermodynamics parameters.

This paper is organized as follows: in the next section we present the model itself and discuss some properties regarding the discrete symmetries. In section III we calculate the thermodynamics parameters, such as pressure, energy, specific heat and charge density. We also calculate the critical temperature for the Bose-Einstein condensation. In section IV we present our final remarks. 

\section{Model}
The model we are considering consists of the complex scalar sector of the Lorentz-violating extension of the standard model, recently proposed by Kostelecky \cite{Edwards:2018lsn}. The lagrangian describing the system is

\begin{eqnarray}\label{lagrangian1}
\nonumber\mathcal{L}&=&G^{\mu\nu}(\partial_{\mu}\phi)^*\partial_{\nu}\phi-m^2\phi^*\phi-\frac{i}{2}[\phi^*\hat{k}_a^{\mu}\partial_{\mu}\phi-\phi\hat{k}_a^{\mu}(\partial_{\mu}\phi)^*]
\end{eqnarray}
where the tensor $G^{\mu\nu}=g^{\mu\nu}+(\hat{k}_c)^{\mu\nu}$ is composed by the Minkowski metric tensor $g^{\mu\nu}=diag(1,-1,-1,-1)$ and a Lorentz-violating constant tensor $(\hat{k}_c)^{\mu\nu}$. The tensor $(\hat{k}_c)^{\mu\nu}$ and the vector $\hat{k}_a^{\mu}$ violates the Lorentz invariance by breaking the equivalence between particle and observer transformations. Such tensors, assumed to be constant, imply the independence of the space-time position, which yields translational invariance assuring the conservation of momentum and energy. Note that while the tensor $(\hat{k}_c)^{\mu\nu}$ is dimensionless, the vector $\hat{k}_a^{\mu}$ has dimension of mass.

Regarding the analysis of the discrete symmetries on $\hat{k}_a^{\mu}$ and $(\hat{k}_c)^{\mu\nu}$, the results are summarized in the table. As we can see the vector $\hat{k}_a^{\mu}$ is CPT-odd while $(\hat{k}_c)^{\mu\nu}$ is CPT-even. The PT symmetry is always preserved, however effects of CP violation can be saw with the $\hat{k}_a^{0}$ and $(\hat{k}_c)^{0i}$ components. It is important to highlight here that the $\hat{k}_a^i$ violates the symmetries of charge conjugation, parity inversion and time reversal simultaneously.

\begin{table}[ht!]
\begin{tabular}{|l|l|l|l|l|}
\hline
 & C & P & T & CPT \\ \hline
\,\,\,\,\,\,\,\,\,\,\,\,$\hat{k}_a^{0}$ & - & +  & +  & \,\,\,\,- \\ \hline
\,\,\,\,\,\,\,\,\,\,\,\,$\hat{k}_a^{i}$ & - & -  & -  & \,\,\,\,- \\ \hline
$(\hat{k}_c)^{00}$, $(\hat{k}_c)^{ij}$ & + & + & + & \,\,\,+ \\ \hline
\,\,\,\,\,\,\,\,\,\,$(\hat{k}_c)^{0i}$ & + & - & - & \,\,\,+ \\ \hline
\end{tabular}
\end{table}

\section{Finite Temperature Effects}

For our purposes we will consider only the contributions from $\hat{k}_a^{\mu}$, since the influence of $(\hat{k}_c)^{\mu\nu}$ was already addressed in \cite{Casana:2011bv}. The lagrangian (\ref{lagrangian1}) possess an obvious $U(1)$ symmetry, so that
\begin{equation}
    \phi\rightarrow \phi'=e^{-i\alpha}\phi,
\end{equation}
with $\alpha\in\mathbb{R}$. The Noether's theorem states that for any given continuous symmetry there is a conserved quantity in connection. In order to find such conserved quantity let us consider $\alpha$ as a function of space-time position $\alpha=\alpha(x)$, then
\begin{eqnarray}
\mathcal{L}'=\mathcal{L}+\phi^*\phi\partial_{\mu}\alpha\partial^{\mu}\alpha +i\partial_{\mu}\alpha(\phi^*\partial^{\mu}\alpha-\phi\partial^{\mu}\phi^*)-\phi^*\phi\hat{k}_a^{\mu}\partial_{\mu}\alpha.
\end{eqnarray}
The Euler-Lagrange equation gives us the equation of motion for the ``field'' $\alpha(x)$. The contribution $\partial\mathcal{L}/\partial\alpha=0$, so that,
\begin{equation}
    \frac{\partial\mathcal{L}}{\partial(\partial_{\nu}\alpha)}=\phi^*\phi(2\partial^{\nu}\alpha-\hat{k}_a^{\nu})+i(\phi^*\partial^{\nu}\phi-\phi\partial^{\nu}\phi^*)
\end{equation}
is a conserved quantity. Letting $\alpha(x)$ be a constant again we obtain the following conserved current,
\begin{equation}
    j^{\nu}(x)=i(\phi^*\partial^{\nu}\phi-\phi\partial^{\nu}\phi^*)-\phi^*\phi\hat{k}_a^{\nu}.
\end{equation}
Making use of the equations of motion for $\phi$ and $\phi^*$, given by
\begin{eqnarray}
-\Box\phi^*-m^2\phi^*+i\hat{k}_a^{\mu}\partial_{\mu}\phi^*&=&0\\
-\Box\phi-m^2\phi-i\hat{k}_a^{\mu}\partial_{\mu}\phi&=&0,
\end{eqnarray}
a direct calculation show us the conservation of the four current, i.e., $\partial_{\mu}j^{\mu}(x)=0$. Consequently the charge density is expressed by
\begin{eqnarray}
    \nonumber Q&=&\int d^3x j^0\\
    &=&\int d^3x\left[i\left(\phi^*\frac{\partial\phi}{\partial t}-\phi\frac{\partial\phi^*}{\partial t}\right)-\phi^*\phi\hat{k}_a^0\right].
\end{eqnarray}
It is convenient to split the fieds $\phi$ and $\phi^*$ into two real components $\phi_1$ and $\phi_2$ as
\begin{eqnarray}
    \phi&=&\frac{1}{\sqrt{2}}(\phi_1+i\phi_2)\\
    \phi^*&=&\frac{1}{\sqrt{2}}(\phi_1-i\phi_2),
\end{eqnarray}
so that the Lagrangian can be rewritten in terms of $\phi_i$ with $i=1,2$ as follows
\begin{eqnarray}
    \mathcal{L}&=&\frac{1}{2}\partial_{\mu}\phi_i\partial^{\mu}\phi_i-\frac{m^2}{2}\phi_i\phi_i+\frac{1}{2}\phi_i\epsilon_{ij}\hat{k}_a^{\mu}\partial_{\mu}\phi_j.
\end{eqnarray}
The canonically conjugated momenta are:
\begin{eqnarray}
    \pi_i=\frac{\partial\phi_i}{\partial t}+\frac{1}{2}\epsilon_{ij}\phi_j\hat{k}_a^{0}
\end{eqnarray}
then the Hamiltonian becomes
\begin{eqnarray}
    \nonumber\mathcal{H}&=&\frac{1}{2}\left(\pi_i\pi_i+(\vec{\nabla}\phi_i)\cdot(\vec{\nabla}\phi_i)+\left(m^2+\frac{1}{4}(\hat{k}_a^0)^2\right)\phi_i\phi_i\right.\\
    &&\left.-\phi_i\epsilon_{ij}\vec{k}_a\cdot\vec{\nabla}\phi_j-\phi_i\epsilon_{ij}\hat{k}_a^0\pi_j\right).
\end{eqnarray}
The charge density can also be expressed in terms of $\phi_i$, 
\begin{equation}
    Q=\int d^3x\epsilon_{ij}\pi_i\phi_j.
\end{equation}
Letting $\mathcal{H}(\phi,\pi)\rightarrow\mathcal{H}(\phi,\pi)-\mu\mathcal{N}(\phi,\pi)$, being $\mu$ the chemical potential and $\mathcal{N}(\phi,\pi)$ the conserved charge density, identified as $Q$, the partition function becomes:
\begin{eqnarray}\label{Z1}
    \nonumber Z&=&\int D\pi_i\int_{periodic}D\phi_i\exp\left\{\int_0^{\beta}d\tau\int d^3x\times\right.\\
    &&\times\left.\left[i\pi_i\frac{\partial\phi_i}{\partial\tau}-\mathcal{H}(\phi_i,\pi_i)+\mu\epsilon_{ij}\pi_i\phi_j\right]\right\}.
\end{eqnarray}
The term ``periodic'' means that the integration over the field is constrained in such way that $\phi(\vec{x},0)=\phi(\vec{x},\beta)$ with $\beta=1/T$. The partition function can be written as
\begin{eqnarray}\label{Z2}
    \nonumber Z&=&\int D\pi_i\int_{periodic}D\phi_i\exp\left\{\int_0^{\beta}d\tau\int d^3x\times\right.\\
    \nonumber&&\times\left.\left[-\frac{1}{2}\pi_i^2+\left(\frac{1}{2}\phi_i\epsilon_{ij}\hat{k}_a^0+i\frac{\partial\phi_j}{\partial\tau}-\mu\epsilon_{ij}\phi_j\right)\pi_j\right.\right.\\
    &&\left.\left.-\frac{1}{2}(\vec{\nabla}\phi_i)^2-\frac{1}{2}\left(m^2+\frac{1}{4}(\hat{k}_a^0)^2\right)\phi_i^2+\phi_i\epsilon_{ij}\vec{k}_a\cdot\vec{\nabla}\phi_j\right]\right\}.
\end{eqnarray}

So that the integration over the momenta can be straightforwardly done. Then we obtain,
\begin{eqnarray}\label{Z3}
    \nonumber Z&=&(N')^2\int_{periodic}D\phi_i\exp\left\{\int_0^{\beta}d\tau\int d^3x\times\right.\\
    \nonumber&&\times\left.\left[\frac{1}{2}\left(i\frac{\partial\phi_j}{\partial\tau}-\mu\epsilon_{ij}\phi_i\right)^2+\frac{i}{2}\phi_i\epsilon_{ij}\phi_j\hat{k}_a^0-\frac{\mu}{2}\phi_i^2\hat{k}_a^0\right.\right.\\
    &&\left.\left.-\frac{1}{2}(\vec{\nabla}\phi_i)^2-\frac{1}{2}m^2\phi_i^2+\phi_i\epsilon_{ij}\vec{k}_a\cdot\vec{\nabla}\phi_j\right]\right\}.
\end{eqnarray}
The factor $N'$ is a normalization constant, irrelevant in the present context, since multiplication of $Z$ by any constant will not change the thermodynamics. The components of $\phi$ can be Fourier-expanded as,
\begin{eqnarray}\label{Fourier}
    \phi_1&=&\sqrt{2}\zeta\cos\theta+\sqrt{\frac{\beta}{V}}\sum_{n}\sum_{\vec{p}}e^{i(\vec{p}\cdot\vec{x}+\omega_n\tau)}\phi_{1;n}(\vec{p})\\
    \phi_2&=&\sqrt{2}\zeta\sin\theta+\sqrt{\frac{\beta}{V}}\sum_{n}\sum_{\vec{p}}e^{i(\vec{p}\cdot\vec{x}+\omega_n\tau)}\phi_{2;n}(\vec{p}),
\end{eqnarray}
where $\omega_n=2\pi n T$, owing to the constraint of periodicity that $\phi(\vec{x},\beta)=\phi(\vec{x},0)$ for all $\vec{x}$. Here $\zeta$ and $\theta$ are independent of $(\vec{x},\tau)$ and determine the full infrared behaviour of the field; that is, $\phi_{1;0}(\vec{p}=\vec{0})=\phi_{1;0}(\vec{p}=\vec{0})=0$. This allows for the possibility of condensation of the bosons into the zero-momentum state. Substituting (\ref{Fourier}) into (\ref{Z2}) the partition function becomes
\begin{equation}
    Z=(N')^2\prod_n\prod_p\int D\phi_{1;n}(\vec{p})D\phi_{2;n}(\vec{p})e^{S},
\end{equation}
where $S$ is given by

\begin{eqnarray}
    \nonumber S&=&\beta V(\mu^2+\mu\hat{k}_a^0-m^2)\zeta^2\\
    &&-\frac{1}{2}\sum_n\sum_p\left(\phi_{1;-n}(-\vec{p}),\phi_{2;-n}(-\vec{p})\right)D\left(\begin{array}{c}
\phi_{1;n}(\vec{p})   \\
\phi_{1;n}(\vec{p})
    \end{array}\right),
\end{eqnarray}
being $D$

\begin{eqnarray}
    D=\beta^2\left(\begin{array}{cc}
    \omega_n^2+\omega^2-\mu^2+\mu\hat{k}_a^0 & -2\mu\omega_n+\omega_n\hat{k}_a^0+i\vec{k}_a\cdot\vec{p} \\
    2\mu\omega_n-\omega_n\hat{k}_a^0-i\vec{k}_a\cdot\vec{p} & \omega_n^2+\omega^2-\mu^2+\mu\hat{k}_a^0
    \end{array}\right),
\end{eqnarray}
with $\omega=\sqrt{\vec{p}^2+m^2}$. Carrying out the integrations over $\phi_{1;n}$ and $\phi_{2;n}$, we have,
\begin{equation}
    \ln Z=\beta V(\mu^2+\mu\hat{k}_a^0-m^2)\zeta^2+\ln (\det D)^{-1/2},
\end{equation}
so that we can rewritte in the following form

\begin{eqnarray}\label{generatingfunctional}
    \nonumber\ln Z&=&\beta V(\mu^2+\mu\hat{k}_a^0-m^2)\zeta^2-\frac{1}{4}V\int\frac{d^3p}{(2\pi)^3}\left\{\beta(\Sigma+\Lambda)\right.\\
    \nonumber&&+2\ln\left[1-e^{-\frac{1}{2}\beta(\Sigma-\hat{k}_a^0+2\mu)}\right]+2\ln\left[1-e^{-\frac{1}{2}\beta(\Sigma+\hat{k}_a^0-2\mu)}\right]\\
    \nonumber&&\left.+2\ln\left[1-e^{-\frac{1}{2}\beta(\Lambda-\hat{k}_a^0+2\mu)}\right]+2\ln\left[1-e^{-\frac{1}{2}\beta(\Lambda+\hat{k}_a^0-2\mu)}\right]\right\},
\end{eqnarray}
with

\begin{eqnarray}
    \Sigma&=&\sqrt{-4\vec{k}_a\cdot\vec{p}+(\hat{k}_a^0)^2+4\omega^2}\\
    \Lambda&=&\sqrt{4\vec{k}_a\cdot\vec{p}+(\hat{k}_a^0)^2+4\omega^2}.
\end{eqnarray}
It is important to point out that the above expression for $\ln Z$ was obtained under the consideration of a convergence condition which states that
\begin{equation}\label{convergence}
    \left|\mu-\frac{\hat{k}_a^0}{2}\right|\leq m.
\end{equation}
In the absence of the Lorentz violation parameter we recover the usual convergence condition ($|\mu|\leq m$) already known in the literature, first stated in the work of Haber \cite{Haber:1981fg}. 
The usual relation 
\begin{equation}
    \frac{PV}{T}=\ln Z,
\end{equation}
gives us the equation of state for the system. The pressure can be calculated as
\begin{eqnarray}\label{pressurecomplete}
    \nonumber P&=&\frac{1}{\beta}\frac{\partial}{\partial V}\ln Z\\
    \nonumber&=&(\mu^2+\mu\hat{k}_a^0-m^2)\zeta^2-\frac{1}{4\beta}\int\frac{d^3p}{(2\pi)^3}\left\{\beta(\Sigma+\Lambda)\right.\\
    \nonumber&&+2\ln\left[1-e^{-\frac{1}{2}\beta(\Sigma-\hat{k}_a^0+2\mu)}\right]+2\ln\left[1-e^{-\frac{1}{2}\beta(\Sigma+\hat{k}_a^0-2\mu)}\right]\\
    \nonumber&&\left.+2\ln\left[1-e^{-\frac{1}{2}\beta(\Lambda-\hat{k}_a^0+2\mu)}\right]+2\ln\left[1-e^{-\frac{1}{2}\beta(\Lambda+\hat{k}_a^0-2\mu)}\right]\right\},
\end{eqnarray}

The internal energy can be written as follows:
\begin{eqnarray}
    \nonumber E&=&-\frac{\partial}{\partial \beta}\ln Z\\
   \nonumber &=&-V(\mu^2+\mu\hat{k}_a^0-m^2)\zeta^2-\frac{1}{4}V\int\frac{d^3p}{(2\pi)^3}\left\{-\Lambda -\Sigma\right.\\
   \nonumber &&\left.+\frac{-\Sigma+\hat{k}_a^0-2\mu}{e^{\frac{1}{2}\beta(\Sigma-\hat{k}_a^0+2\mu)}-1}-\frac{\Sigma+\hat{k}_a^0-2\mu}{e^{\frac{1}{2}\beta(\Sigma+\hat{k}_a^0-2\mu)}-1}\right.\\
   &&\left.+\frac{-\Lambda+\hat{k}_a^0-2\mu}{e^{\frac{1}{2}\beta(\Lambda-\hat{k}_a^0+2\mu)}-1}-\frac{\Lambda+\hat{k}_a^0-2\mu}{e^{\frac{1}{2}\beta(\Lambda+\hat{k}_a^0-2\mu)}-1}\right\}.
\end{eqnarray}
The specific heat at constant volume can be expressed by

\begin{eqnarray}
    \nonumber C_v&=&\frac{\partial E}{\partial T}\\
    \nonumber&=&\frac{1}{32}V\beta^2\int\frac{d^3p}{(2\pi)^3}\left\{\left(\hat{k}_a^0-2\mu+\Sigma\right)^2\csch^2\left[\frac{1}{4}\beta\left(\hat{k}_a^0-2\mu+\Sigma\right)\right]\right.\\
    \nonumber &&+\left(-\hat{k}_a^0+2\mu+\Sigma\right)^2\csch^2\left[\frac{1}{4}\beta\left(-\hat{k}_a^0+2\mu+\Sigma\right)\right]\\
    \nonumber &&+\left(-\hat{k}_a^0+2\mu+\Lambda\right)^2\csch^2\left[\frac{1}{4}\beta\left(-\hat{k}_a^0+2\mu+\Lambda\right)\right]\\
    &&\left.+\left(\hat{k}_a^0-2\mu+\Lambda\right)^2\csch^2\left[\frac{1}{4}\beta\left(\hat{k}_a^0-2\mu+\Lambda\right)\right]\right\}.
\end{eqnarray}
The charge density is given by

\begin{eqnarray}
    \nonumber\rho&=&\frac{1}{\beta V}\frac{\partial \ln Z}{\partial \mu}\\
    \nonumber &=& \frac{1}{2}\int\frac{d^3p}{(2\pi)^3}\left[\frac{1}{e^{\frac{1}{2}\beta(\Sigma+\hat{k}_a^0-2\mu)}-1}-\frac{1}{e^{\frac{1}{2}\beta(\Sigma-\hat{k}_a^0+2\mu)}-1}\right.\\
    \nonumber &&\left.+\frac{1}{e^{\frac{1}{2}\beta(\Lambda+\hat{k}_a^0-2\mu)}-1}-\frac{1}{e^{\frac{1}{2}\beta(\Lambda-\hat{k}_a^0+2\mu)}-1}\right]
\end{eqnarray}

At this point is more convenient to work with the contributions from $\hat{k}_a^0$ and $\vec{k}_a$ separately.

%%%%%%%%%%%%%%%%%%%%%%%%%%%%%%%%%%%%%%%%%%%%%%%%%%%%%%%%%%%%%%%%%%%%%%%%%%%%%%%%%%%%%%%%%%%%%%%%%%%%%%%%%%%%%%%%%%%%%%%%%%%%%%%%%%%%%%%%%%%%%%%%%%%%%%%%%%%%%%%%%%%%%%%%%%%%%%%%%%%%%%%%%%%%%%%%%%%%%%%%%%%%%%%%%%%%%%%%%%%%%%%%%%%%%%%%%%%%%%%%%%%%%%%%%%%%%%%%%%%%%%%%%%%%%%%%%%%%%%%%%%%%%%%%%%%%%%%%%%%%%%%%%%%%%%%%%%%%%%%%%%%%%%%%%%%%%%%%%%%%%%%%%%%%%%%%%%%%%%%%%%%%%%%%%%%%%%%%%%%%%%%%%%%%%%%%%%%%%%%%%%%%%%%%%%%%%%%%%%%%%%%%%%%%%%%%%%%%%%%%%%%%%%%

\subsection{$\hat{k}_a^0$ contribution}

We will now analyze the contributions emergent only from $\hat{k}_a^0$. In this case we have:
\begin{equation}\label{SigmaK0}
    \Sigma=\Lambda=\sqrt{(\hat{k}_a^0)^2+4\omega^2}.
\end{equation}
Initially let us analyze the pressure in a non condensate fase, i.e., $\zeta=0$. The expression (\ref{pressurecomplete}) for the pressure simplifies to
\begin{eqnarray}\label{pressurek0}
    \nonumber P&=&-\frac{1}{4\beta}\int\frac{d^3p}{(2\pi)^3}\left\{2\beta\Sigma+4\ln\left[1-e^{-\frac{1}{2}\beta(\Sigma+\hat{k}_a^0-2\mu)}\right]\right.\\
    &&\left.+4\ln\left[1-e^{-\frac{1}{2}\beta(\Sigma-\hat{k}_a^0+2\mu)}\right]\right\}.
\end{eqnarray}
Using spherical coordinates, the integration measure goes to
\begin{equation}
    \int\frac{d^3p}{(2\pi)^3}\longrightarrow\frac{1}{2\pi^2}\int dp p^2.
\end{equation}
So that we can rewrite the pressure as
\begin{eqnarray}\label{pressurek01}
    \nonumber P&=&-\frac{1}{8\pi^2\beta}\int dp p^2\left\{2\beta\Sigma+4\ln\left[1-e^{-\frac{1}{2}\beta(\Sigma+\hat{k}_a^0-2\mu)}\right]\right.\\
    &&\left.+4\ln\left[1-e^{-\frac{1}{2}\beta(\Sigma-\hat{k}_a^0+2\mu)}\right]\right\}.
\end{eqnarray}
An analogous procedure gives us, for the internal energy,
\begin{eqnarray}\label{energyk0}
   \nonumber E &=&-\frac{1}{4\pi^2}V\int dp p^2\times\\
    &&\times\left\{-\Sigma+\frac{-\Sigma+\hat{k}_a^0-2\mu}{e^{\frac{1}{2}\beta(\Sigma-\hat{k}_a^0+2\mu)}-1}-\frac{\Sigma+\hat{k}_a^0-2\mu}{e^{\frac{1}{2}\beta(\Sigma+\hat{k}_a^0-2\mu)}-1}\right\}
\end{eqnarray}
 and for the specific heat,
\begin{eqnarray}\label{Cv0}
    \nonumber C_v &=&\frac{1}{32\pi^2}V\beta^2\int dp p^2\times\\
    \nonumber &&\times \left\{\left(\hat{k}_a^0-2\mu+\Sigma\right)^2\csch^2\left[\frac{1}{4}\beta\left(\hat{k}_a^0-2\mu+\Sigma\right)\right]\right.\\
    &&\left.+\left(-\hat{k}_a^0+2\mu+\Sigma\right)^2\csch^2\left[\frac{1}{4}\beta\left(-\hat{k}_a^0+2\mu+\Sigma\right)\right]\right\}.
\end{eqnarray}

\begin{figure}[ht!]
    \centering
    \includegraphics[scale=0.6]{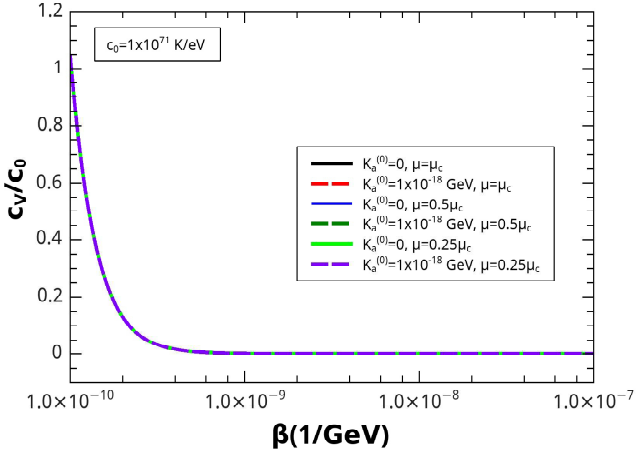}
    \caption{Specific heat}
    \label{fig2}
\end{figure}
The figure (\ref{fig2}) shows the behaviour of the specific heat as a function of $\beta$. Without lost of generality we consider the mass as being the mass of the Higgs boson, i.e., $m_H=125GeV$. As we can see, the specific heat goes to zero when $\beta$ assumes any value grater than $10^{-9}GeV$ and in the limit when $\beta\rightarrow0 (T\rightarrow\infty)$ we have $C_v\rightarrow1$. Also there is no significant difference when we consider $\hat{k}_a^0=10^{-18}GeV$ or $\hat{k}_a^0=0GeV$.

The charge density is given by
\begin{eqnarray}\label{chargedensityk0}
    \rho &=& \frac{1}{2\pi^2}\int dp p^2\left[\frac{1}{e^{\frac{1}{2}\beta(\Sigma+\hat{k}_a^0-2\mu)}-1}-\frac{1}{e^{\frac{1}{2}\beta(\Sigma-\hat{k}_a^0+2\mu)}-1}\right]
\end{eqnarray}

\begin{figure}[ht!]
    \centering
    \includegraphics[scale=0.6]{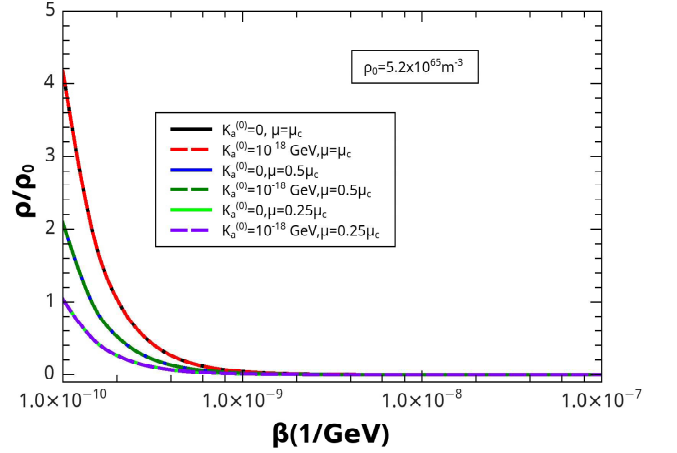}
    \caption{Charge density}
    \label{fig1}
\end{figure}

The figure (\ref{fig1}) presents the behaviour of the charge density for three values of chemical potential, namely, $\mu_c$, $(1/2)\mu_c$ and $(1/3)\mu_c$ being $\mu_c=m+(\hat{k}_a^0/2)$ the critical chemical potential. Note that the plot has two values for $\hat{k}_a^0$, one of them being $\hat{k}_a^0=10^{-18}GeV$ which we set due to the table of bounds present in \cite{Edwards:2019lfb}, and the other value is $\hat{k}_a^0=0GeV$, in order to compare the effect of the Lorentz-violating model with the usual SM. As we can see, there is no considerable effect of Lorenz violation in the behaviour of the charge density. The great difference of sixteen orders of magnitude between the LV parameter and the usual scale of the SM does not provide any observable effect in the charge density.

At this point it is important to discuss the influence of the Lorentz violation parameters in the critical temperature $T_c$ above which we can always find a chemical potential satisfying the convergence condition (\ref{convergence}). For the Bose-Einstein condensation to occur we must have $\mu=\pm m+\frac{\hat{k}_a^0}{2}$. Setting such condition into (\ref{chargedensityk0}), we obtain a correction for the critical temperature, first presented in \cite{Haber:1981fg}, as

\begin{equation}
    T_c=\sqrt{\frac{24|\rho|}{(8-3(\hat{k}_a^0)^2)m}}.
\end{equation}
Note that the correction appears as a second order contribution in the LV parameter. In fact in the perturbative series all odd contributions in the LV parameter are zero, being the second order contribution the dominant term. However due to the extremely small bound for the LV parameter, no significant change in the critical temperature appears.

%%%%%%%%%%%%%%%%%%%%%%%%%%%%%%%%%%%%%%%%%%%%%%%%%%%%%%%%%%%%%%%%%%%%%%%%%%%%%%%%%%%%%%%%%%%%%%%%%%%%%%%%%%%%%%%%%%%%%%%%%%%%%%%%%%%%%%%%%%%%%%%%%%%%%%%%%%%%%%%%%%%%%%%%%%%%%%%%%%%%%%%%%%%%%%%%%%%%%%%%%%%%%%%%%%%%%%%%%%%%%%%%%%%%%%%%%%%%%%%%%%%%%%%%%%%%%%%%%%%%%%%%%%%%%%%%%%%%%%%%%%%%%%%%%%%%%%%%%%%%%%%%%%%%%%%%%%%%%%%%%%%%%%%%%%%%%%%%%%%%%%%%%%%%%%%%%%%%%%%%%%%%%%%%%%%%%%%%%%%%%%%%%%%%%%%%%%%%%%%%%%%%%%%%%%%%%%%%%%%%%%%%%%%%%%%%%%%%%%%%%%%%%%%

\subsection{$\vec{k}_a$ contribution}

In this section we are going to analyze the contributions emergent from the spatial component of the Lorentz-violating vector $(\hat{k}_a)^\mu$ by setting $\hat{k}_a^0=0$. Knowing that the equation (\ref{generatingfunctional}) is invariant under the interchange $\Sigma\leftrightarrow\Lambda$, we are interested in the maximum contribution of the Lorentz-violating parameter, so that without lost of generality we will consider $\vec{k}_a$ and $\vec{p}$ collinear by setting $\theta=0$ into $\vec{k}_a\cdot\vec{p}=|\vec{k}_a||\vec{p}|\cos\theta$. Such simplification yields for the pressure

\begin{eqnarray}\label{pressurek}
    \nonumber P &=&-\frac{1}{4\pi^2\beta}\int dp p^2\left\{\beta(\Sigma+\Lambda)\right.\\
    \nonumber&&+\ln\left[1-e^{-\beta(\Sigma+\mu)}\right]+\ln\left[1-e^{-\beta(\Sigma-\mu)}\right]\\
    &&\left.+\ln\left[1-e^{-\beta(\Lambda+\mu)}\right]+\ln\left[1-e^{-\beta(\Lambda-\mu)}\right]\right\},
\end{eqnarray}
where
\begin{eqnarray}
    \Sigma&=&\sqrt{- k_a p + \omega^2}\\
    \Lambda&=&\sqrt{ k_a p + \omega^2},
\end{eqnarray}
with $k_a=|\vec{k}_a|$. The internal energy becomes

\begin{eqnarray}
   \nonumber E &=&-\frac{1}{4\pi^2}V\int dp p^2 \left\{-\Lambda -\Sigma\right.\\
   \nonumber &&\left.+\frac{-\Sigma-\mu}{e^{\beta(\Sigma+\mu)}-1}-\frac{\Sigma-\mu}{e^{\beta(\Sigma-\mu)}-1}\right.\\
   &&\left.+\frac{-\Lambda-\mu}{e^{\beta(\Lambda+\mu)}-1}-\frac{\Lambda-\mu}{e^{\beta(\Lambda-\mu)}-1}\right\}.
\end{eqnarray}
For the specific heat we obtain
\begin{eqnarray}
    \nonumber C_v &=&\frac{1}{16\pi^2}V\beta^2\int dp p^2\left\{\left(-\mu+\Sigma\right)^2\csch^2\left[\frac{1}{2}\beta\left(-\mu+\Sigma\right)\right]\right.\\
    \nonumber &&+\left(+\mu+\Sigma\right)^2\csch^2\left[\frac{1}{2}\beta\left(+\mu+\Sigma\right)\right]\\
    \nonumber &&+\left(+\mu+\Lambda\right)^2\csch^2\left[\frac{1}{2}\beta\left(+\mu+\Lambda\right)\right]\\
    &&\left.+\left(-\mu+\Lambda\right)^2\csch^2\left[\frac{1}{2}\beta\left(-\mu+\Lambda\right)\right]\right\}.
\end{eqnarray}
Finally, the charge density is presented as
\begin{eqnarray}
    \nonumber \rho &=& \frac{1}{4\pi^2}\int dp p^2\left[\frac{1}{e^{\beta(\Sigma-\mu)}-1}-\frac{1}{e^{\beta(\Sigma+\mu)}-1}\right.\\
    \nonumber &&\left.+\frac{1}{e^{\beta(\Lambda-\mu)}-1}-\frac{1}{e^{\beta(\Lambda+\mu)}-1}\right]
\end{eqnarray}

The behaviour of the charge density and specific heat considering the contributions emergent from $\hat{k}_a^i$ are essentially the same as the ones described in figures (\ref{fig1}) and (\ref{fig2}), respectively. 

\section{Final Remarks}

In this paper we study the corrections emergent from a Lorentz-violating CPT-odd extension of the complex scalar sector to the Bose-Einstein condensation and to the thermodynamic parameters. We initially discussed some features of the model to only then compute the corrections to the Bose-Einstein condensation. The calculations were done by computing the generating functional, from which we extract the thermodynamics parameters. 

We considered only the contributions from the Lorentz-violating vector $\hat{k}_a^{\mu}$, since the influence of $(\hat{k}_c)^{\mu\nu}$ was already addressed in \cite{Casana:2011bv}. We split the vector $\hat{k}_a^{\mu}$ into its spatial and temporal components and treat them separately. 

We shown that considering the contribution from $\hat{k}_a^0$, the converge criteria for the generating functional states that $|\mu-\hat{k}_a^0/2|\leq m$, from which we obtain a Lorentz-violating correction for the critical temperature $T_c$ that sets the Bose-Einstein Condensation. Besides, we also obtained analytical expressions for the pressure, energy, specific heat and charge density for both $\hat{k}_a^0$ and $\hat{k}_a^i$.

A natural continuation of this work consists in the consideration of self interacting scalar fields as well as the computation of the finite temperature effects on the radiative corrections for the Lorentz-violating scalar QED.

\vspace{1cm}
\begin{center}
    {\bf Acknowledgment}
\end{center}

The authors would like to thank prof. J. A. Hela\"{y}el-Neto and the referees for useful suggestions.

\end{document}